\documentclass[prb,twocolumn,amsmath,showpacs]{revtex4}

\usepackage{graphicx}
\usepackage{color}
\usepackage{dcolumn}

\begin{document}

\newcommand{\ket}    [1]{{|#1\rangle}}
\newcommand{\bra}    [1]{{\langle#1|}}
\newcommand{\braket} [2]{{\langle#1|#2\rangle}}
\newcommand{\bracket}[3]{{\langle#1|#2|#3\rangle}}

\def\pw{^{({\rm W})}}
\def\ph{^{({\rm H})}}
\def\k{{\bf k}}
\def\R{{\bf R}}
\def\b{{\bf b}}
\def\q{{\bf q}}
\def\o{{\cal O}}
\def\e{{\cal E}}
\def\v{{\rm v}}
\def\pw{^{({\rm W})}}
\def\ph{^{({\rm H})}}
\def\la{\langle\kern-2.0pt\langle}
\def\ra{\rangle\kern-2.0pt\rangle}
\def\vt{\vert\kern-1.0pt\vert}
\def\D{{D}\ph}
\def\n{{\cal N}}
\def\u{{\cal U}}

\newcommand{\red}{\textcolor{red}}
\newcommand{\blue}{\textcolor{blue}}
\newcommand{\green}{\textcolor{green}}
\newcommand{\cyan}{\textcolor{cyan}}
\newcommand{\magenta}{\textcolor{magenta}}

\newcommand{\sihao}{\fontsize{14pt}{\baselineskip}\selectfont}      
\newcommand{\xiaosihao}{\fontsize{12pt}{\baselineskip}\selectfont}  
\newcommand{\wuhao}{\fontsize{10.5pt}{\baselineskip}\selectfont}    
\newcommand{\xiaowuhao}{\fontsize{9pt}{\baselineskip}\selectfont}   
\newcommand{\liuhao}{\fontsize{7.875pt}{\baselineskip}\selectfont}  
\newcommand{\qihao}{\fontsize{5.25pt}{\baselineskip}\selectfont}    

\title{Electronic Structures and Surface States of Topological
  Insulator Bi$_{1-x}$Sb$_{x}$}

\author{Hai-Jun Zhang$^1$, Chao-Xing Liu$^2$, Xiao-Liang Qi$^3$, Xiao-Yu Deng$^1$,\\
 Xi Dai$^1$, Shou-Cheng Zhang$^3$ and Zhong Fang$^1$}

\affiliation{$^1$ Beijing National Laboratory for Condensed Matter
  Physics, and Institute of Physics, Chinese Academy of Sciences,
  Beijing 100190, China;}

\affiliation{$^3$ Center for Advanced Study, Tsinghua
  University,Beijing, 100084, China}

\affiliation{$^2$ Department of Physics, McCullough Building, Stanford
  University, Stanford, CA 94305-4045;}

\date{\today}

\begin{abstract}
  We investigate the electronic structures of the alloyed Bi$_{1-x}$Sb$_x$
  compounds based on first-principle calculations including
  spin-orbit coupling (SOC), and calculate the surface states of
  semi-infinite systems using maximally localized Wannier function
  (MLWF). From the calculated results, we analyze the topological
  nature of Bi$_{1-x}$Sb$_x$, and found the followings: (1) pure Bi
  crystal is topologically trivial; (2) topologically non-trivial phase can be
  realized by reducing the strength of SOC  via Sb doping; (3) the
  indirect bulk band gap, which is crucial to realize the true bulk
  insulating phase, can be enhanced by uniaxial pressure along $c$
  axis. (4) The calculated surface states can be compared with
  experimental results,  which confirms the topological nature; (5) We
  predict the spin-resolved Fermi surfaces and showed the vortex
  structures, which should be examined by future experiments.
\end{abstract}

 \maketitle

\section{Introduction}
\label{sec:intro}

In an ordinary insulator, the valence and conduction bands are
separated by an energy gap, making it electrically inert. Therefore,
the ordinary insulator is not sensitive to the change of boundary
condition. Recently a new class of insulator, namely topological
insulator (TI), is proposed~\cite{kane2005B,
bernevig2006a,konig2008,fu2007A,moore2007,qi2008B}. TI also has a
bulk energy gap, which is usually generated by spin-orbit coupling
(SOC); however it is different from the ordinary insulator in the
sense that topologically protected gapless states, robust against
disorder, appear at the edge or surface of a finite sample within
the bulk energy gap. Thus the TI has conducting channels along its
edge or surface. The quantum spin Hall (QSH) insulator, such as
HgTe/CdTe quantum well~\cite{bernevig2006a,konig2007a,konig2008}, is
an example of two-dimensional (2D) TI. The conducting edge channels
of HgTe/CdTe quantum wells have been theoretically
predicted\cite{bernevig2006a} and experimentally
observed\cite{konig2007a,konig2008}. From the theoretical point of
view, the TI can be distinguished from the ordinary insulator by the
$Z_2$ topological
invariants~\cite{kane2005B,fu2007B,moore2007,qi2008B}, and the
existence of gapless spin-filtered edge states on the sample
boundary is guaranteed for TI. The edge states come in Kramers's
doublets, and time reversal (TR) symmetry ensures the crossing of
their energy band at time reversal invariant momenta (TRIM). Since
these band crossings on the edge are protected by TR, they can not
be removed by any perturbation respecting the TR symmetry, such as
non-magnetic impurities. It is expected that the robust gapless
spin-filtered surface(edge) states have novel applications in
spintronics.

Besides the 2D QSH insulator, the TI can also exist in three
dimensional (3D)
material\cite{fu2007A,fu2007B,dai2008,hsieh2008,teo2008,zhang2008,xia2008}.
Similar to the edge states in 2D QSH insulator, in 3D TI,
topological surface state protected by TR, which can be described by
odd number of Dirac points , emerges at the surface of the finite 3D
sample\cite{fu2007A,fu2007B,qi2008B,liu2008}. Compared with the 2D
TI, the 3D TI and its surfaces can be readily investigated by ARPES
and STM experiments. The 3D TI also displays the remarkable
topological magneto-electric effect\cite{qi2008B}. Therefore
searching for realistic 3D TI is now becoming an attractive and
challenging subject.

It was first suggested that the semiconducting alloy of bismuth and
antimony (Bi$_{1-x}$Sb$_{x}$) is an example of such 3D
TI\cite{fu2007B,teo2008}. Based on the tight-binding (TB) model of
Liu and Allen\cite{liu1995}, Fu and Kane developed a theory to
analyze the topological nature of the surface state in
Bi$_{1-x}$Sb$_{x}$ alloy~\cite{teo2008}.  Experimentally, Hsieh {\it
et al}\cite{hsieh2008} observed the surface states by
high-momentum-resolution angle-resolved photoemission spectroscopy
(ARPES), and demonstrated the topological nature of the surface
states by counting the number of the Fermi surface crossings from
the Zone center to the boundary. However, clear discrepancies exist
between the theory and the experiment about the surface states,
although their final conclusions are consistent with each other.  On
the other side, the surface states of pure Bi or Sb have been
intensely studied experimentally and
theoretically\cite{sugawara2006,hofmann2006,ast2003A,koroteev2004,ast2003B,hirahara2006},
but there are still fewer careful studies of their alloy. Therefore,
in this paper, we present a systematic study of the surface states
of Bi$_{1-x}$Sb$_{x}$ alloy, based on quantitative first principle
calculations. We show that pure Bi is topologically trivial because
the SOC is too strong. The effective role of Sb doping is to reduce
the strength of SOC and revert the band ordering at L point of the
Brillouin Zone(BZ). Finally strong topological insulator can be
reached by Sb doping.  By constructing the maximally localized
Wannier function (MLWF) from the {\it ab-initio} schemes, we
calculate the surface states of semi-infinite system, and analyze
the shape of Fermi surfaces as well as the spin-resolved local
density of states. These results are compared with the experiment of
Hsieh {\it et al}\cite{hsieh2008}. In addition, although direct band
gap exists in Bi$_{1-x}$Sb$_{x}$ system, indirect band gap can be
only realized for a very narrow doping range, and the material has
long been regarded as typical semi-metallic system. In order to make
the bulk material insulating completely, based on our calculations,
we predict that an efficient way to enhance the indirect band gap is
to apply uniaxial pressure along the $c$-axis.

The paper is organized as follows. In sec.~\ref{sec:struct} we
discuss the crystal structure and symmetry of Bi$_{1-x}$Sb$_{x}$. In
Sec.~\ref{sec:TBmodel} we study the transition between the
topological non-trivial and trivial phases, and present a schematic
phase diagram as a function of SOC.  In Sec.~\ref{sec:FPmethod} we
develop an accurate method based on MLWF to obtain the surface Green
Function, and study the topologically non-trivial surface states of
Bi$_{1-x}$Sb$_{x}$. In Sec.~\ref{sec:conclusion} we provide a brief
discussion and conclusion.

\section{Structure and Symmetry}
\label{sec:struct}

\begin{figure}
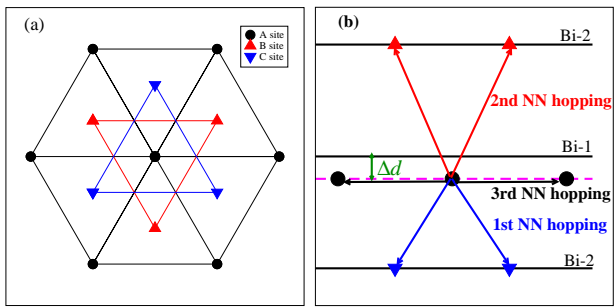

  \includegraphics[clip,width=4cm]{fig1a}
   \includegraphics[clip,width=4cm,height=4cm]{fig1b}
   \caption{ (color online) (a) Along the (111) direction of
     rhombohedral A7 structure, there are three possible atomic
     positions(A,B,C) for the triangle plane. Black filled dots denote A
     site, red filled up-triangles denote B site, and blue filled
     down-triangles denote C site.  (b) The schematic plot of Bi-layers
     projected onto the plane paralleling to the $(111)$ axis. After
     the dimerization of two Bi layers, the Bi-2 moves to the dashed
     site. $\Delta d$ denotes the magnitude of the dimerization. 1st and
     2nd nearest neighbor(NN) hopping are the inter-layer hopping
     between the different sublattices, while 3rd NN hopping is the
     intra-layer hopping within the same sublattice.}\label{fig:hex}
\end{figure}

Bi and Sb have the same rhombohedral A7 crystal
structure\cite{liu1995} with space group $R\bar{3}m$. The A7 structure
can be regarded as a distorted fcc NaCl structure. For fcc NaCl
structure, there are two sets of sublattice, say Bi-1 for Na sites and
Bi-2 for Cl sites, which both form the fcc structure. In such fcc
structure, Bi-1 and Bi-2 sublattices are equivalent and we can shift
Bi-1 sublattice by (1/2,0,0), or (0,1/2,0), or (0,0,1/2) of cubic
structure to obtain Bi-2 sublattice. Along (111) direction of fcc
structure, Bi triangle layers are stacked with the sequence of
$ABCABCABC\cdots$, where A, B and C denote three different atomic
positions for triangle plane, as shown in Fig.\ref{fig:hex}. Without
distortions, there exist two kinds of space-inversion center. One is
located at the Bi layer center and each Bi sublattice is
space-inverted to itself (called type-I inversion), while the other
one is located at the middle way of two Bi layers and each Bi
sublattice is space-inverted to another sublattice (called type-II
inversion).

\begin{figure}
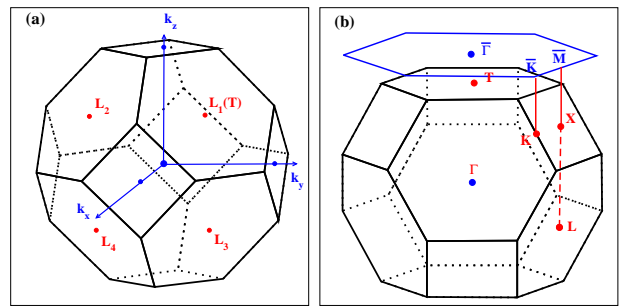

  \includegraphics[clip,width=4cm]{fig2a}
   \includegraphics[clip,width=4cm]{fig2b}
   \caption{(color online) (a) BZ of fcc structure.  (b) 3D BZ of
     rhombohedral A7 structure and its projection onto the [111]
     surface. The A7 BZ can be obtained from the fcc BZ by two steps:
     (1) rotating the (111) direction of cubic to be along the
     $c$-axis; (2) slightly distortion of the BZ along the
     $c$-axis. The $L_1$ point in fcc BZ is changed to be T point in
     the A7 BZ, which is now inequivalent to $L_{2,3,4}$ due to the
     distortion. T ($\Gamma$), X (L), and K in 3D BZ are projected to
     $\overline{\Gamma}$, $\overline{M}$ and $\overline{K}$ in the 2D
     BZ of [111] surface.}
\label{fig:BZ}
\end{figure}

Starting from fcc NaCl structure, two steps are required to obtain
rhombohedral A7 structure. One is the stretching along cubic (111)
direction, while the second is the relative shift of inter-Bi-layer
distances along (111) direction, or in other words dimerization of
two Bi layers.  The second step breaks the type-I inversion
symmetry, but preserve the type-II inversion. Therefore, after
distortions, the inversion symmetry can only transfer the atoms from
one sublattice to the other, which is very important for our
following discussion.

The Brillouin Zone(BZ) of fcc structure and rhombohedral A7
structure are shown in Fig.\ref{fig:BZ}. For fcc structure, there
are four equivalent L points, which are located at
$L_1$=($\pi$,$\pi$,$\pi$), $L_2$=($\pi$,$-\pi$,$\pi$),
$L_3$=($\pi$,$\pi$,$-\pi$) and $L_4$=($\pi$,$-\pi$,$-\pi$) of BZ.
After two kinds of distortions along cubic (111) direction, the fcc
structure changes to the rhombohedral A7 structure, which breaks the
equivalence between $L_1$ (which is denoted as T point in BZ of
rhombohedral A7 structure) and $L_{2,3,4}$.

\section{Effect of Alloying and Phase Diagram} \label{sec:TBmodel}

Although pure Bi and Sb have been studied extensively, the alloyed
system is not carefully considered yet. In this section, we will
present a simple TB model by taking the strength of SOC $\lambda$
and dimerization $\Delta d$ as two key parameters to describe the
effect of alloying. Based on this TB model, we obtained a schematic
phase diagram for the topological nature of the compounds, which is
instructive for us to understand the main physics. Then in the next
section, surface states of alloyed system will be studied from
accurate {\it ab-initio} calculations based on virtual crystal
approximation (VCA).

\subsection{Effect of Alloying} \label{sec:TB_model}

In order to take into account the effect of alloying, certain kinds
of approximation have to be introduced. The first step, which is
conventionally followed, is to assume the uniform distribution and
neglect the disorder effects. However, this approximation is not
sufficient. To further simplify our understanding, we emphasize the
following factors: (1) Sb is located just on top of Bi in the
periodical table, therefore they have the same number of valence
electrons and form the same A7 crystal structure. (2) Even 30\% Sb
alloying into Bi will only modify the lattice parameters by around
1\%~\cite{liu1995}. (3) The atomic SOC strength of Sb is weaker than
Bi by a factor of 3. Therefore, we believe the strongest effect of
Sb alloying into Bi is to reduce the SOC strength, and we can
neglect the effect coming from the change of lattice parameters.
Following this strategy, we construct a simple TB model to
understand the main physics.

We consider one $s$ and three $p$ orbitals of each Bi atoms,
together with the two sublattices and two spin degree of freedom,
and totally there are 16 orbitals, denoted as
$|s_{1}^\uparrow\rangle$,$|s_{1}^\downarrow\rangle$,
$|p_{1x}^\uparrow\rangle$, $|p_{1y}^\uparrow\rangle$,
$|p_{1z}^\uparrow\rangle,$ $|p_{1x}^\downarrow\rangle$,
$|p_{1y}^\downarrow\rangle$, $|p_{1z}^\downarrow\rangle,$
$|s_{2}^\uparrow\rangle$,$|s_{2}^\downarrow\rangle$,
$|p_{2x}^\uparrow\rangle$, $|p_{2y}^\uparrow\rangle$,
$|p_{2z}^\uparrow\rangle,$ $|p_{2x}^\downarrow\rangle$,
$|p_{2y}^\downarrow\rangle$, $|p_{2z}^\downarrow\rangle$.  Here $z$
axis is taken along the (111) direction of fcc structure or the
(001) direction of hexagonal cell and subscript number denotes
different sublattice.  The hopping parameters are defined in
Fig.\ref{fig:hex}, with the expression
\begin{equation}
\begin{split}
  V_{pp\sigma}  &= \widetilde{V}_{pp\sigma}(1+\alpha (\Delta d-\Delta d_0) )\\
  V_{pp\pi}     &= \widetilde{V}_{pp\pi}(1+\beta (\Delta d-\Delta d_0) )\\
  V_{ss\sigma}  &= \widetilde{V}_{ss\sigma}(1+\gamma (\Delta d-\Delta d_0) )\\
  V_{sp\sigma}  &= \widetilde{V}_{sp\sigma}(1+\delta (\Delta d-\Delta d_0) )
 \end{split}
\end{equation}
for the first nearest-neighbour (NN) hopping,
\begin{equation}
\begin{split}
  V_{pp\sigma}' &= \widetilde{V}_{pp\sigma}'(1-\alpha (\Delta d-\Delta d_0) )\\
  V_{pp\pi}'    &= \widetilde{V}_{pp\pi}'(1-\beta (\Delta d-\Delta d_0) )\\
  V_{ss\sigma}' &= \widetilde{V}_{ss\sigma}'(1-\gamma (\Delta d-\Delta d_0) )\\
  V_{sp\sigma}' &= \widetilde{V}_{sp\sigma}'(1-\delta (\Delta d-\Delta d_0) )
 \end{split}
\end{equation}
for the second NN hopping and
\begin{equation}
\begin{split}
  V_{pp\sigma}''  &= \widetilde{V}_{pp\sigma}''\\
  V_{pp\pi}''     &= \widetilde{V}_{pp\pi}''\\
  V_{ss\sigma}''  &= \widetilde{V}_{ss\sigma}''\\
  V_{sp\sigma}''  &= \widetilde{V}_{sp\sigma}''
 \end{split}
\end{equation}
for the third NN hopping.  Here the parameters $\tilde{V}_m$,
$\tilde{V}'_m$ and $\tilde{V}''_m$ are taken from the Liu-Allen
model\cite{liu1995}, where $m$ is $pp\sigma$, $pp\pi$, $ss\sigma$ or
$sp\sigma$, respectively.  Since the intra-layer Bi-Bi distance is
larger than inter-layer Bi-Bi distance, $V_m$ and $V'_m$ are larger
than $V''_m$.  Linear dependence of $\Delta d$ due to the
dimerization is assumed and $\Delta d_0$ is for the experiment
structure of pure Bi. When $\Delta d=0$, this system has no
dimerization, leading to $V_m=V'_m$. Then a set of linear equations
is obtained, which can be used to determine the value of the
parameters $\alpha$, $\beta$, $\gamma$ and $\delta$.  Besides the
hopping terms, the atomic SOC with parameter $\lambda$ is also taken
into account.  Therefore, the final Hamiltonian is given as the
function of two variables $\Delta d$ and $\lambda$, with the form
\begin{eqnarray}
\left[
\begin{array}{c|c}
  H_{11}  & H_{12} \\
  & \\
  \hline
  & \\
  H_{21}  & H_{22}
\end{array}
\right]
\end{eqnarray}
where $H_{11}$=$H_{22}$, $H_{12}$=$H_{21}^\dag$ are $8\times8$
matrice. $H_{11}$ includes intra-sublattice hopping and on-site SOC
interaction, while $H_{12}$ represents inter-layer hopping. Since
the Hamiltonian has the type II inversion symmetry at T and L
points, a unitary transformation is applied here to rewrite the
Hamiltonian in the new basis with unambiguous parity,
the odd parity basis,
$\frac{1}{\sqrt{2}}(\ket{s_{1}^\sigma}-\ket{s_{2}^\sigma})$,
$\frac{1}{\sqrt{2}}(\ket{p_{1x}^\sigma}+\ket{p_{2x}^\sigma})$,
$\frac{1}{\sqrt{2}}(\ket{p_{1y}^\sigma}+\ket{p_{2y}^\sigma})$,
$\frac{1}{\sqrt{2}}(\ket{p_{1z}^\sigma}+\ket{p_{2z}^\sigma})$ and
the even parity basis,
$\frac{1}{\sqrt{2}}(\ket{s_{1}^\sigma}+\ket{s_{2}^\sigma})$,
$\frac{1}{\sqrt{2}}(\ket{p_{1x}^\sigma}-\ket{p_{2x}^\sigma})$,
$\frac{1}{\sqrt{2}}(\ket{p_{1y}^\sigma}-\ket{p_{2y}^\sigma})$,
$\frac{1}{\sqrt{2}}(\ket{p_{1z}^\sigma}-\ket{p_{2z}^\sigma})$.
With the new basis, the Hamiltonian is changed to be
$\widetilde{H}$,
\begin{eqnarray}
\left[
\begin{array}{c|c}
 \widetilde{H}_{11}  & 0         \\
    &   \\
   \hline
    &   \\
  0          & \widetilde{H}_{22}
\end{array}
\right]\label{eq:TBtilde}
\end{eqnarray}
which is block diagonal, because the odd and even parity states will
not mix in a system with space-inversion symmetry.

\subsection{Phase Diagram} \label{sec:diagram}

\begin{figure}
\includegraphics[clip,width=4cm,height=4cm]{fig4a.eps}
\includegraphics[clip,width=4cm,height=4cm]{fig4b.eps}\\
\includegraphics[clip,width=4cm,height=4cm]{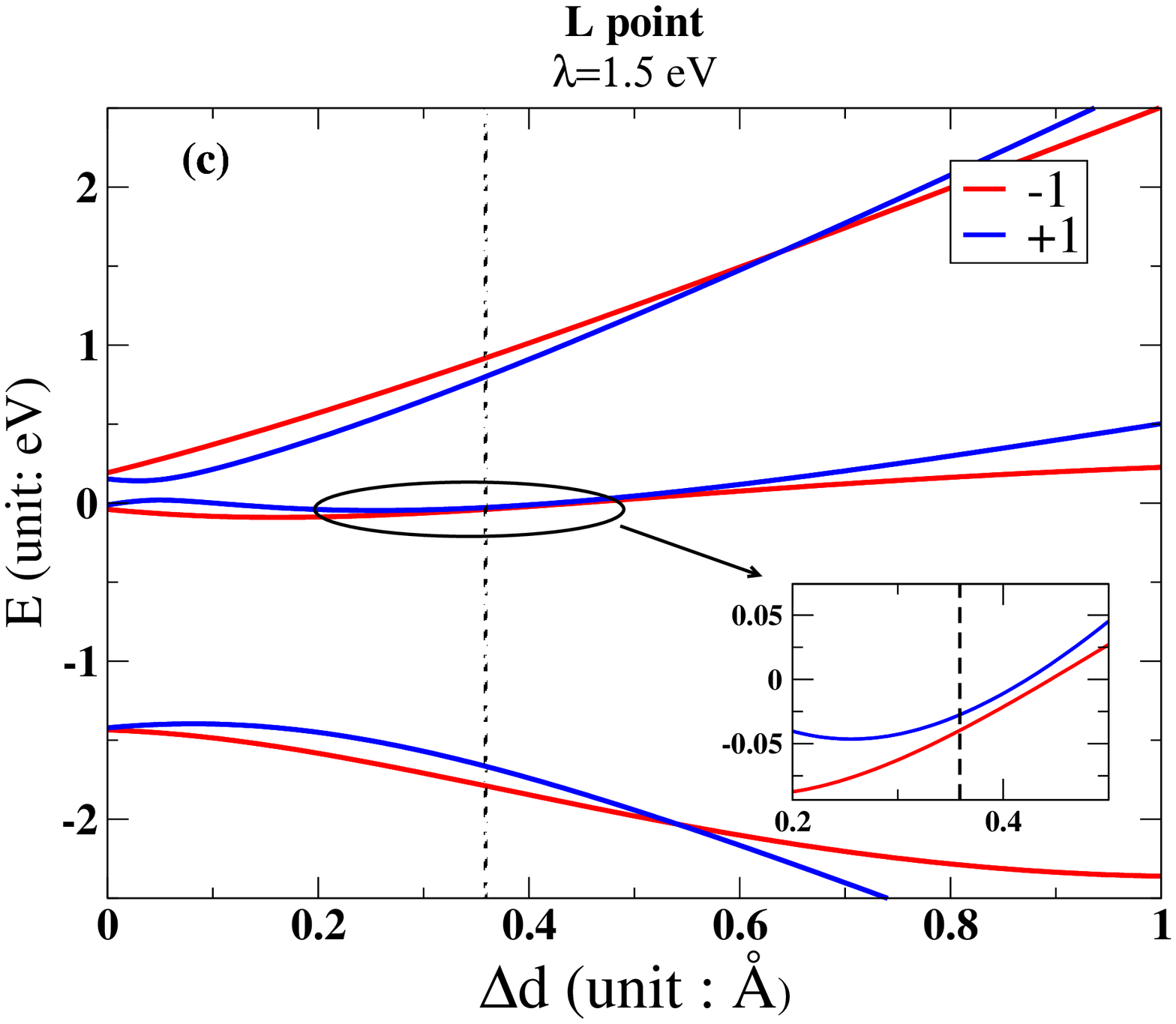}
\includegraphics[clip,width=4cm,height=4cm]{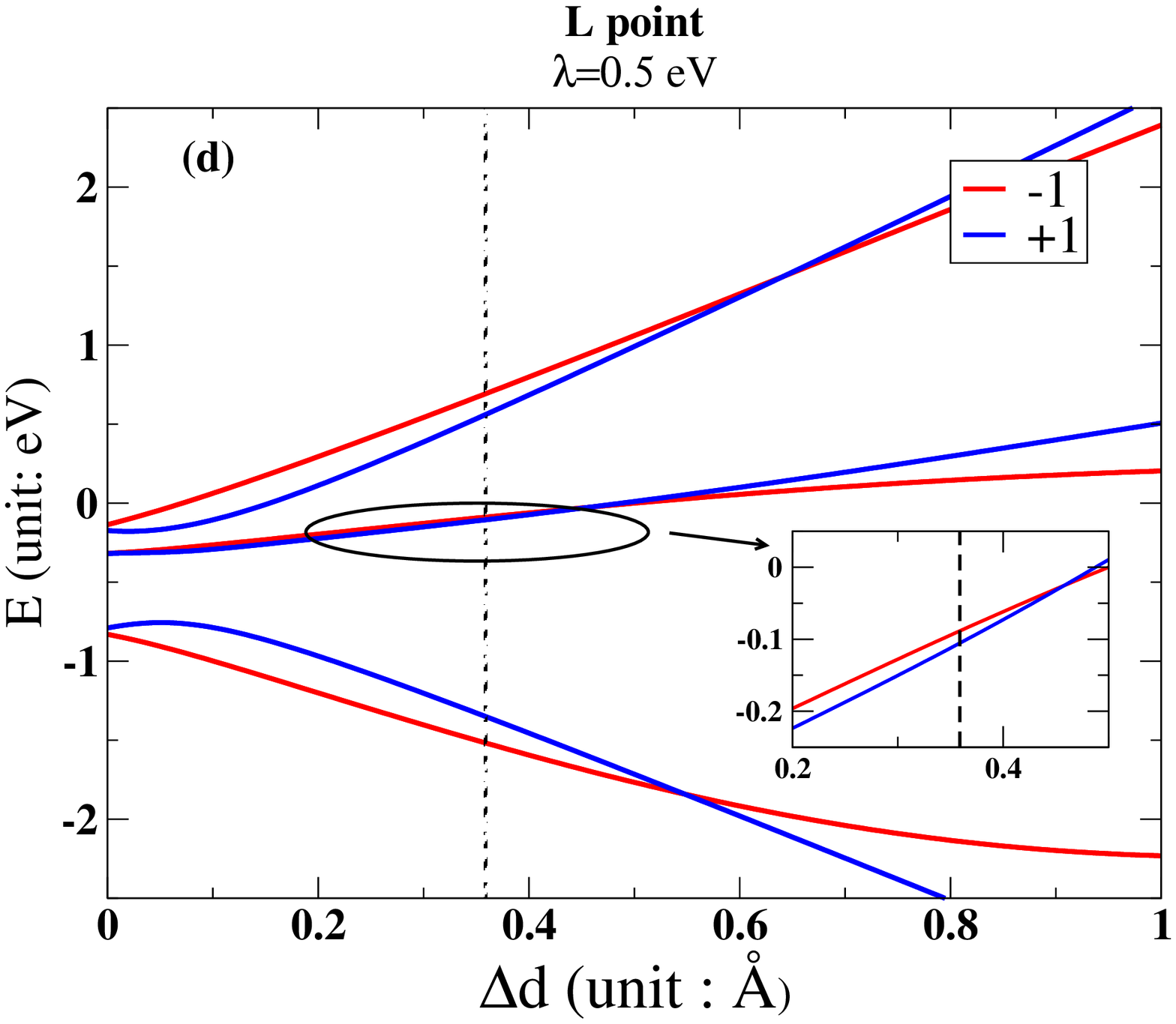}\\
\caption{(color online) Energy bands are plotted as a function of
the dimerization parameter($\Delta d$) with SOC parameter $\lambda$
taken as $1.5eV$ and $0.5eV$. (a) and (b) are for T point, while (c)
and (d) for L point. The blue lines denote the states with parity
$+1$ while the red ones are the states with parity $-1$. The dashed
line represents $\Delta d=\Delta d_0$. }\label{fig:dE}
\end{figure}

The topological nature of the system can be determined from the
parity of the occupied bands at TRIM~\cite{fu2007B}. The band gap
for Bi$_{1-x}$Sb$_x$ is near T and L points, therefore here we focus
on one T and three L points in BZ. The parity of the occupied bands
for T and L points can be easily obtained since the eigen states of
$\tilde{H}_{11}$ have odd parity while those of $\tilde{H}_{22}$
have even parity. In Fig {\ref{fig:dE}}, the energy levels of six
$p$ bands for both L and T points are plotted as a function of
$\Delta d$, where two different values of $\lambda$ are chosen,
$\lambda =1.5$ corresponding to the value of Bi and $\lambda=0.5$
corresponding to the value of Sb. Three lowest levels of total six
$p$ bands should be occupied, namely the conduction band and valence
band are the third lowest band and fourth lowest band respectively,
which have opposite parities. When increasing the dimerization
parameter $\Delta d$, at T point the band gap increases rapidly and
there is no leveling crossing between the conduction band and
valence band, while at L point, the band gap is quite small and the
sequence of conduction band and valence band can even change with
$\lambda=0.5 eV$. When $\Delta d=\Delta d_0$, as indicated by the
dashed line in Fig {\ref{fig:dE}}, the occupied valence bands for Bi
($\lambda=1.5 eV$) and Sb ($\lambda=0.5eV$) have different parities
at L point but the same at T point, therefore we conclude the
topological natures of Bi and Sb are different. This result is the
same as that of Fu and Kane\cite{fu2007B}, in which they claim that
$Z_2$ invariants ($\nu_0;\nu_1\nu_2\nu_3$) are (0;000) for Bi, which
is topologically trivial, but (1;111) for Sb, which corresponds to
topological non-trivial phase.

\begin{figure}
  \includegraphics[width=7.5cm]{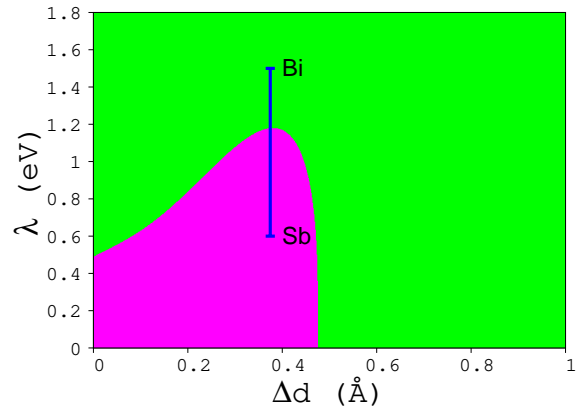}
  \caption{(color online) The phase diagram of the system with two
    variables: the dimerization parameter $\Delta d$ and SOC parameter
    $\lambda$.  In the magenta region, the parity of L and T points
    are +1 and -1, respectively, in other words, the system is in the
    topological non-trivial phase with $Z_2$ invariants (1;111). In
    the green region, however, the parity of L and T points are all
    -1, i.e, it is topological trivial with $Z_2$ invariants
    (0;000). Pure Bi locates in the trivial region and Sb locates at
    topological non-trivial region. The experimental $\Delta d$ of Bi
    and Sb are almost the same~\cite{liu1995}.}
  \label{fig:d_lamda}
\end{figure}

By determining the gap closing line of the conduction and valence
band at L points, we can obtain the phase diagram of the alloy
Bi$_{1-x}$Sb$_{x}$ as a function of SOC parameter $\lambda$ and
dimerization parameter $\Delta d$, as shown in
Fig.~\ref{fig:d_lamda}. Since the topological nature of the system
can only be changed by closing bulk gap, in the magenta region, the
system should have the same non-trivial topological behavior to Sb,
 while in the green region, it should be trivial insulator,
which is same to Bi. It can be also understood from the parity
analysis that the Bi is topologically trivial because both L and T
points have reverted bands ordering, namely parity are -1 for both L
and T. However, by reducing the SOC strength (i.e. Sb doping), the
parity of L is recovered to be +1 while T remains to be -1,
therefore topological non-trivial system is realized.  In the next
section, accurate simulation from {\it ab-initio} calculations will
be presented.

\section{{\it ab-initio} Calculations and Surface States}
\label{sec:FPmethod}

\subsection{{\it ab-initio} Method and Surface Green Function}
\label{sec:ab_green}

The {\it ab-initio} calculation is carried out by our BSTATE
(Beijing Simulation Tool for Atom Technology) \cite{fang2002} code
with plane wave pseudo-potential method. The generalized gradient
approximation(GGA) of PBE-type\cite{perdew1996} is used for the
exchange-correlation potential. Especially in Bi's pseudo-potential,
we take into account $5d^{10}$ electrons as the valence band by
ultra-soft pseudo-potential scheme. The k-mesh is taken as $12\times
12\times12$ and the cut-off energy is $340.0eV$ for the
self-consistent calculation. For pure Bi, the optimized lattice
parameters are $a=4.669{\AA}, c=12.1506\AA$, and $d=0.2341\AA$,
which are in good agreement with previous calculations.

To take into account the effect of alloying, virtual crystal
approximation (VCA) is necessary. There are several ways to do this,
particularly in the pseudo-potential approach, the simple linear
combination of Sb and Bi pseudo-potentials can be used, and the
corresponding pseudo-potential of alloyed virtual atom is
regenerated by solving the atomic problem again~\cite{fang2002}.
Unfortunately, this procedure is not accurate enough for our purpose
here. As already suggested by previous studies~\cite{fang2002}, such
VCA procedure can be used for those states far away from the Fermi
level, however for those states very close to the Fermi level, the
error bar is big. The system we study here (Bi$_{1-x}$Sb$_x$) is
very sensitive to the $p$ orbitals, such VCA pseudo-potential can
not give sufficient accuracy. In order to have an accurate VCA
scheme, we need to consider the particularity of our system. As we
already explained in the last section, the alloyed Bi$_{1-x}$Sb$_x$
have the same crystal structure and almost the same structure
parameters. The main effect of Sb alloying is to tune the SOC
strength, $\lambda$. Therefore we may have a simple yet accurate VCA
scheme. Here we take Bi's parameters for simplicity and tune
$\lambda$ in Bi's pseudo-potential to simulate the doping parameter
$x$ of the alloyed Bi$_{1-x}$Sb$_x$. In such a way, since we do not
need to the solve the atomic problem again, the pseudo-potential is
accurate enough.

We are interested in the surface states of the semi-infinite system,
a method based on maximally localized Wannier function (MLWF) is
developed to calculate the surface states of semi-infinite system.
The {\it ab-initio} MLWF\cite{marzari1997,souza2001} method can be
regarded as an exact TB method with its parameters calculated from
{\it ab-initio} self-consistent electronic structure calculations.
First, the semi-infinite Bi$_{1-x}$Sb$_{x}$ system can be divided
into two parts: the bulk part and the surface part. The bulk part
Hamiltonian is constructed with the MLWFs from bulk
Bi$_{1-x}$Sb$_{x}$ {\it ab-initio} calculations, while the surface
part Hamiltonian is constructed with MLWFs from Bi$_{1-x}$Sb$_{x}$
film with slab calculations. With these MLWFs' hopping parameters,
iterative method \cite{sancho1984,sancho1985} is adopted to solve
the surface Green function of the semi-infinite system
$G^{l\alpha,l\alpha}_{nn}({\k_{\|}}, \epsilon + i\eta)$, where $n$
and $l$ denote the super-cell along z direction and the atomic
bilayer plane within one super-cell respectively. $\alpha$ gives the
orbital index in one atomic bilayer 
and $\k_{\|}$ is a good quantum number in semi-infinite system.

The charge density of states (DOS) and spin DOS\cite{dai2008} are
related to the surface Green function with the expression
\begin{equation}\label{eq:dos}
    N^{l}_{n}({\k_{\|}},\epsilon) =
    -\frac{1}{\pi}\mathbf{Im}\sum_{\alpha}G^{l\alpha,l\alpha}_{nn}({\k_{\|}},\epsilon + i\eta)
\end{equation}
and
\begin{equation}\label{eq:sdos}
\begin{split}
   S^{l}_{n,\sigma}({\k_{\|}},\epsilon) &=
  -\frac{1}{\pi}\mathbf{Im}\sum_{\alpha,\beta}
   G^{l\alpha,l\beta}_{nn}({\k_{\|}},\epsilon +i\eta)O^{\sigma}_{l\beta,l\alpha}\\
   O^{\sigma}_{\beta,\alpha}&=\bracket{l\beta}{\hat{s}^{\sigma}}{l\alpha}
\end{split},
\end{equation}
respectively, where $\hat{s}^{\sigma}$ is the spin($s_{x,y,z}$)
operator. When $n=0$, $l=0$, $N^{0}_{0}({\k_{\|}},\epsilon)$ and
$S^{0}_{0,\sigma}({\k_{\|}},\epsilon)$ give the local DOS and local
spin DOS at the surface and in the following we use
$N({\k_{\|}},\epsilon)$ and $S_\sigma({\k_{\|}},\epsilon)$ for
short.

\subsection{Surface State and Fermi surface }
\label{sec:ESFS}

\begin{figure}
\includegraphics[width=3.5cm,angle=90]{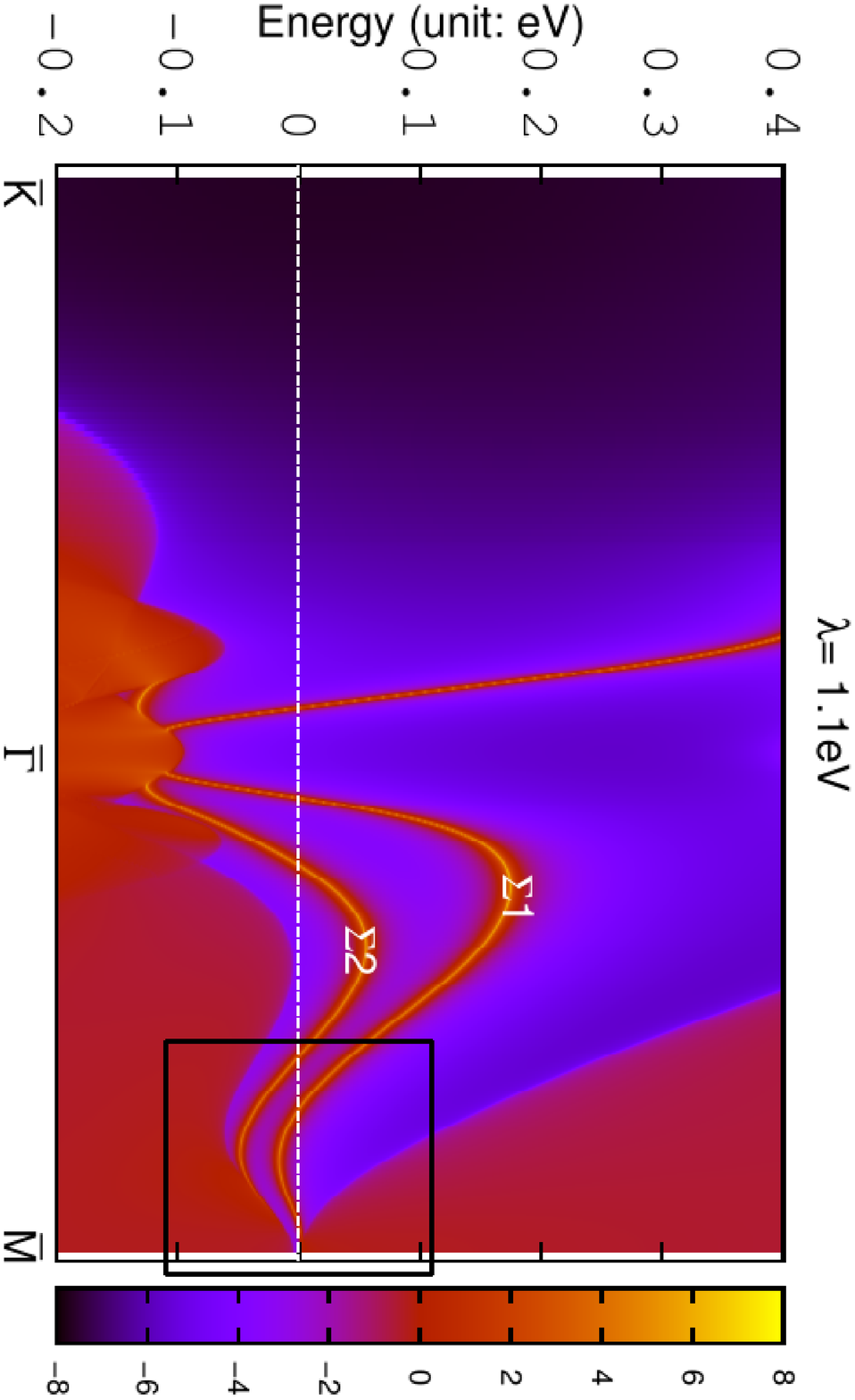}
\includegraphics[width=2.8cm,angle=0]{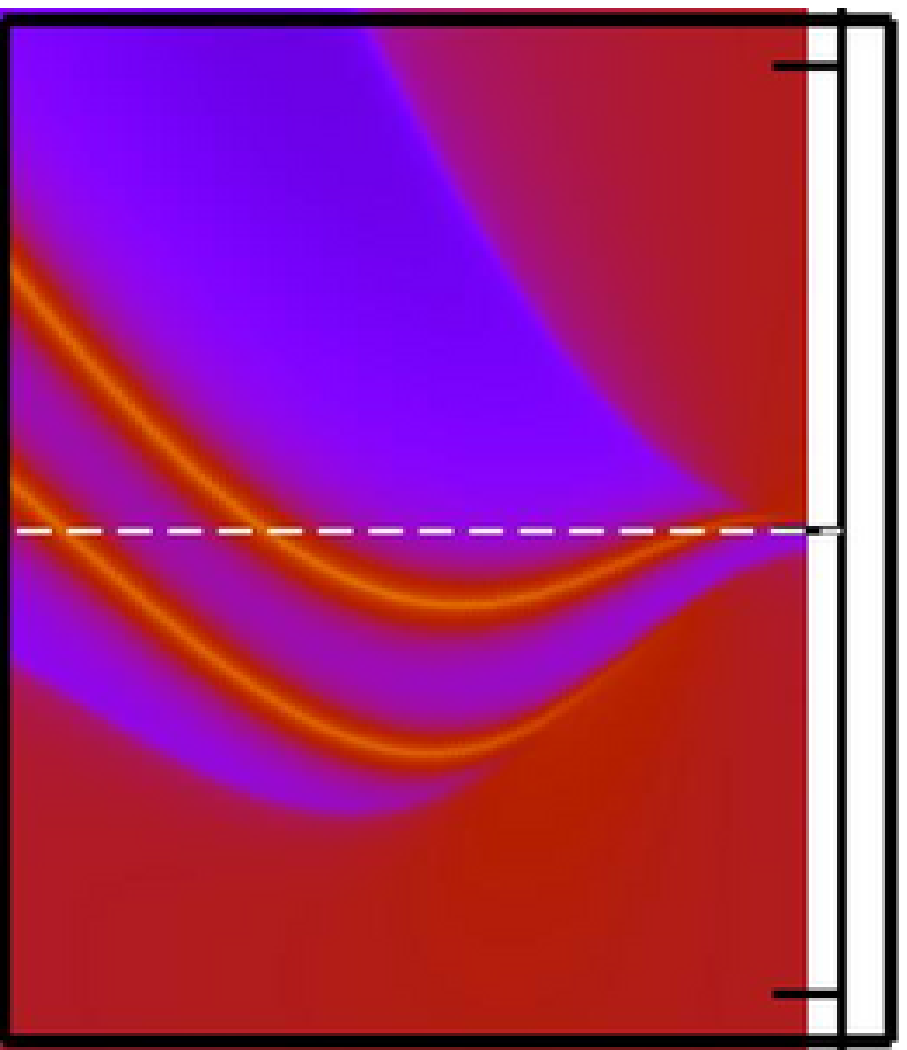}
\caption{(color online) Left panel: The surface local DOS for $\lambda
  = 1.1eV$. The broad red regions denote the continuous bulk bands
  with a small gap of about $10meV$ at $\bar{M}$ point. Two surface
  bands indicated $\Sigma_1$ and $\Sigma_2$ disperse within the bulk
  gap. The dashed line indicates the Fermi energy, which intersects
  five times with the two surface bands from $\bar{\Gamma}$ to
  $\bar{M}$.  Right panel: The region framed by the black rectangle in
  the left panel is zoomed in. One surface state ($\Sigma_1$) goes up
  to merge into the conduction band while the other one ($\Sigma_2$)
  goes back to the valence band.} \label{fig:degecorrect_topo}
\end{figure}

\begin{figure}
\includegraphics[width=3.5cm,angle=90]{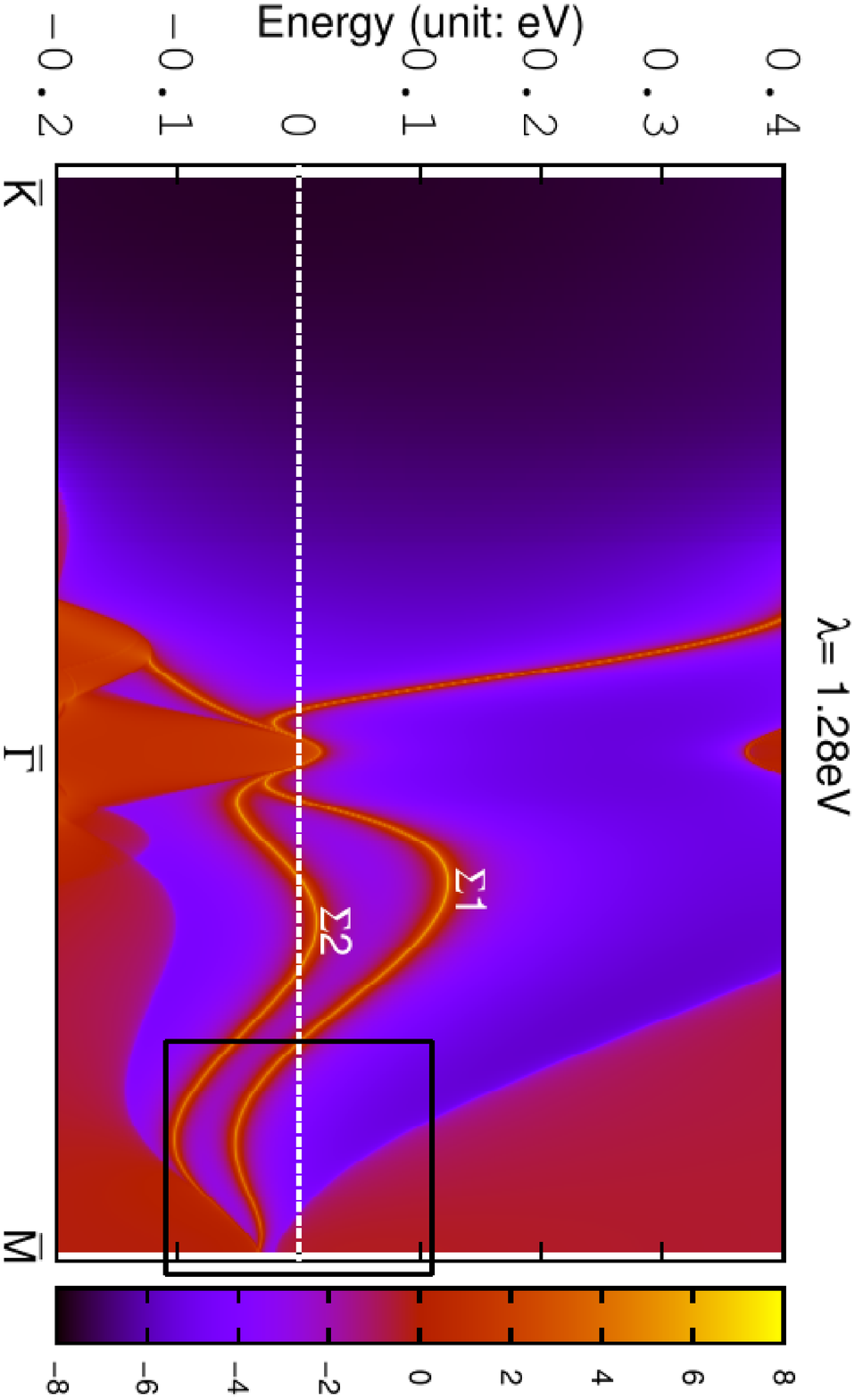}
\includegraphics[width=2.8cm,angle=0]{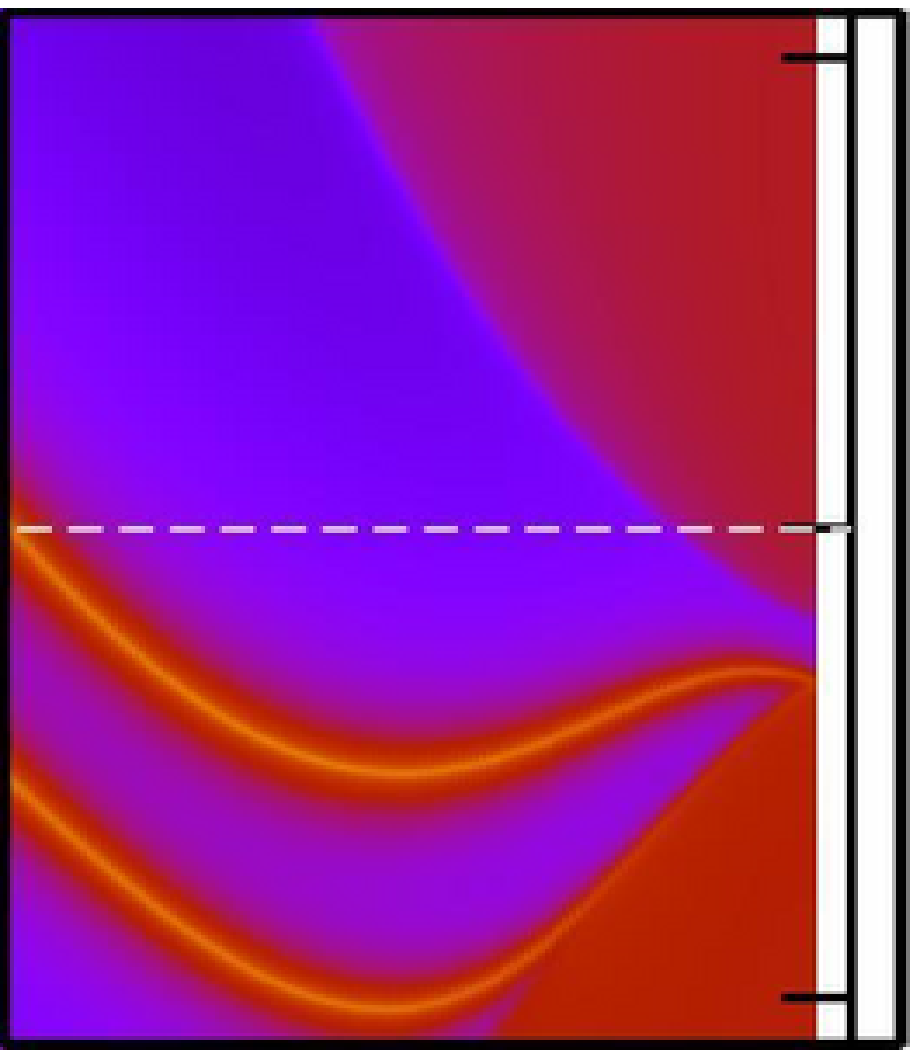}
\caption{(color online) Left panel: The surface local DOS for $\lambda
  = 1.28eV$. The two surface states $\Sigma_1$ and $\Sigma_2$ disperse
  in a different way in the present case. At the $\bar{M}$ point, both
  $\Sigma_1$ and $\Sigma_2$ are connect to the valence band. Therefore
  the Fermi energy intersects four times with the surface bands
  between $\overline{\Gamma}$ and $\overline{M}$. Right panel: The
  region framed by the black rectangle in the left panel is zoomed
  in. } \label{fig:degecorrect_notopo}
\end{figure}

The local DOS $N({\k_{\|}},E)$ at the surface is plotted for two
different SOC parameters, $\lambda$=1.28eV and $\lambda$=1.1eV for
the topologically trivial and non-trivial situations respectively.
In Fig.  \ref{fig:degecorrect_topo} with $\lambda=1.1eV$, developing
from the $\bar{\Gamma}$ point, there exists two surface bands
connected to the $\bar{M}$ point, which are denoted as $\Sigma_1$
and $\Sigma_2$ respectively. At $\bar{M}$ point, $\Sigma_2$ band
returns to valence band while $\Sigma_1$ band merges into the
conduction band. Therefore, those surface states cross the Fermi
energy five times in total (odd number), which indicates the
topologically non-trivial nature of this phase. On the contrary, in
Fig.~\ref{fig:degecorrect_notopo} (for $\lambda$=$1.28eV$), both
$\Sigma_1$ and $\Sigma_2$ bands return to the valence band at
$\bar{M}$ point, and they cross the Fermi level four times (even
number). This indicates that the system is topological trivial.

\begin{figure}
\includegraphics[clip,width=5cm,angle=0]{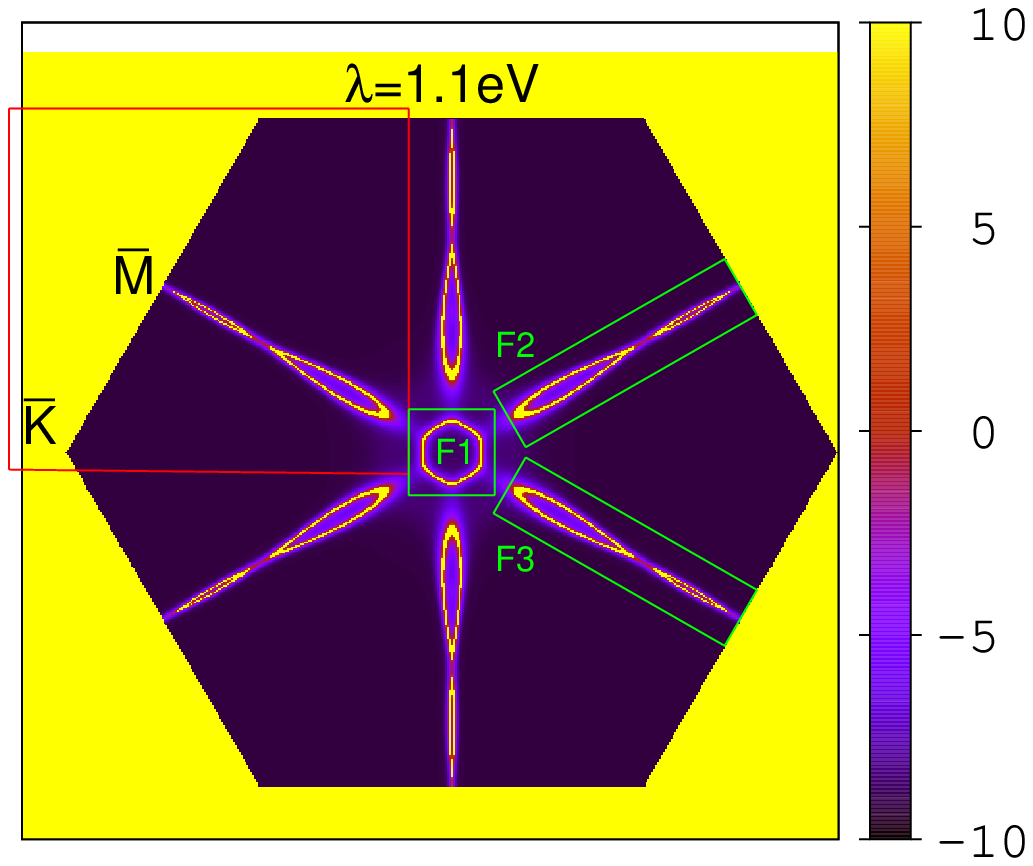}
\includegraphics[clip,width=3.5cm,angle=0]{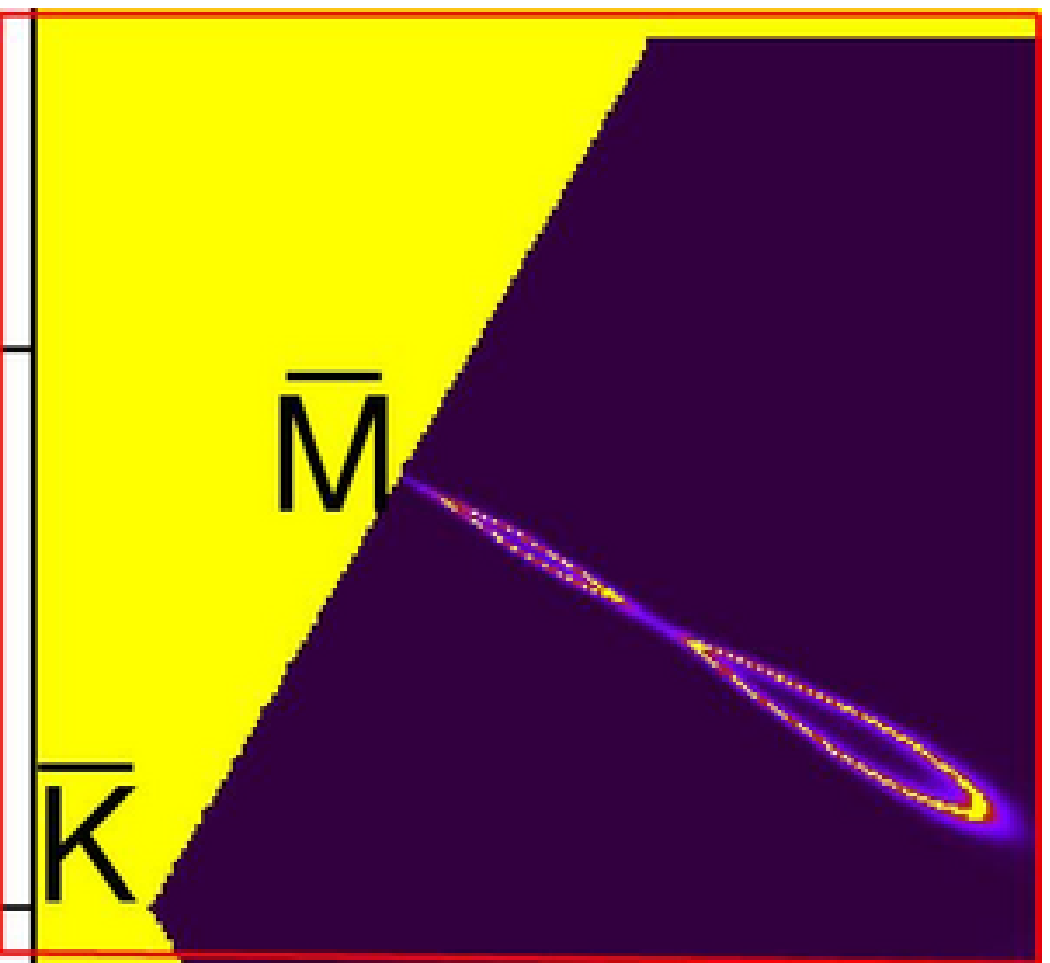}
\caption{(color online) Left panel: The Fermi surface plot for
  $\lambda = 1.1eV$.  The black hexagonal region is the 2D BZ of [111]
  surface for A7 structure.  $\overline{\Gamma}$ is enclosed by a
  hexagonal electron pocket. There are other six hole pockets and six
  electron pockets surrounding.  Right panel: The region framed by red
  rectangle in the left panel is zoomed in. We can clearly see that
  the outest six small electron pocket don't enclose $\overline{M}$.}
 \label{fig:FStotal_topo}
\end{figure}

\begin{figure}
\includegraphics[clip,width=5cm,angle=0]{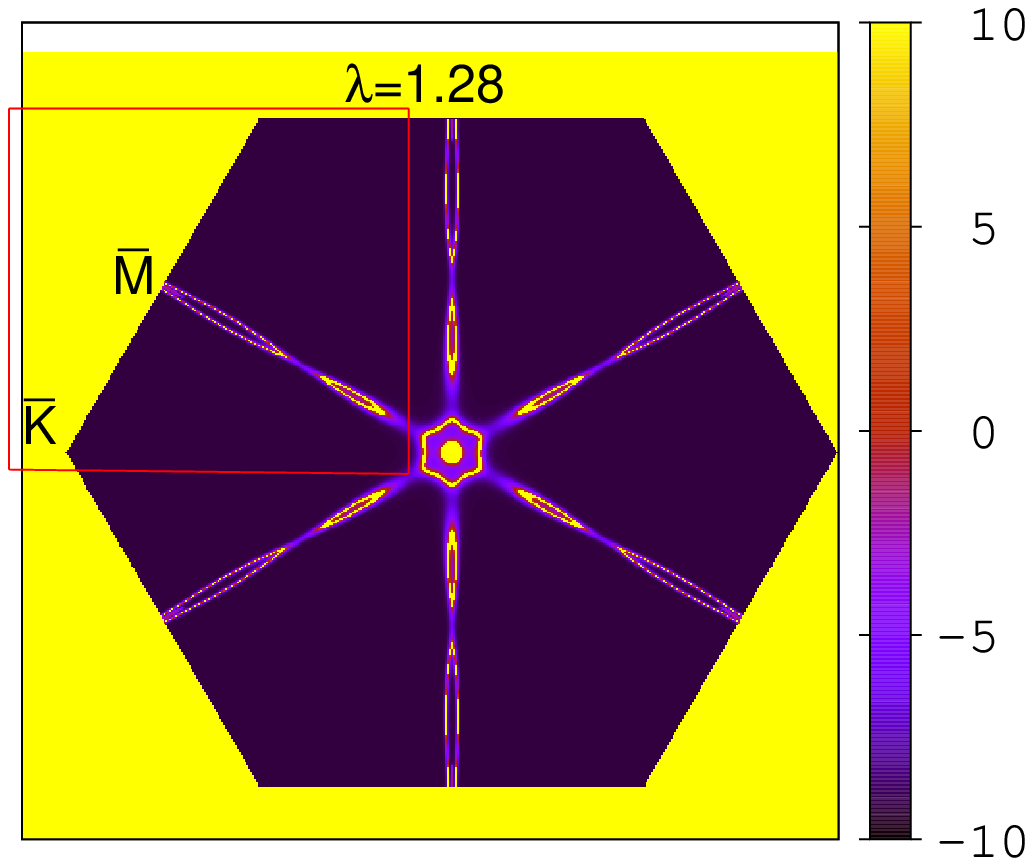}
\includegraphics[clip,width=3.5cm,angle=0]{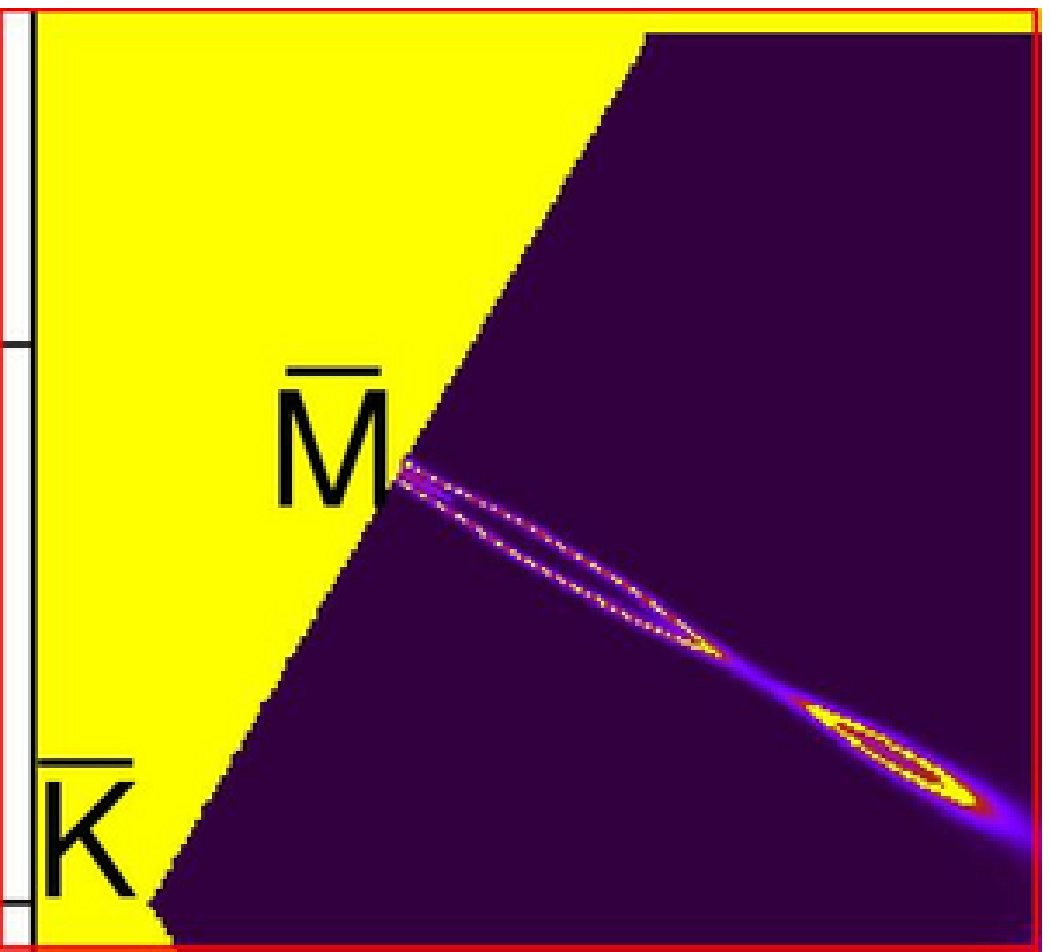}
\caption{(color online) Left panel: The Fermi surface plot for
  $\lambda = 1.28eV$, which is similar to Fig. \ref{fig:FStotal_topo},
  except that the outest six electron pockets enclose the
  $\overline{M}$ point.  Right panel: The region framed by red
  rectangle in the upper panel is zoomed in.
} \label{fig:FStotal_notopo}
\end{figure}

The shape of the Fermi surface for the two different phases is
plotted in Fig.~\ref{fig:FStotal_topo} and
Fig.~\ref{fig:FStotal_notopo}. There are one $\bar{\Gamma}$ point
and three $\bar{M}$ points in the surface BZ, and they are all TRIM.
The main difference between Fig.~\ref{fig:FStotal_topo} and
Fig.~\ref{fig:FStotal_notopo} is around the $\bar{M}$ point.  For
both cases, the $\bar{\Gamma}$ point is enclosed by one Fermi arc,
however, the $\bar{M}$ point is different: it is enclosed by one
Fermi arc at $\lambda=1.28eV$, and it is not for $\lambda=1.1eV$.
Therefore by counting the total number of TRIM enclosed by Fermi
surface in the BZ, it is even number for $\lambda=1.28eV$ (one
$\bar{\Gamma}$ plus three $\bar{M}$), and odd number for
$\lambda=1.1eV$ (only one $\bar{\Gamma}$ point).

\begin{figure}
\includegraphics[clip,width=8cm,angle=0]{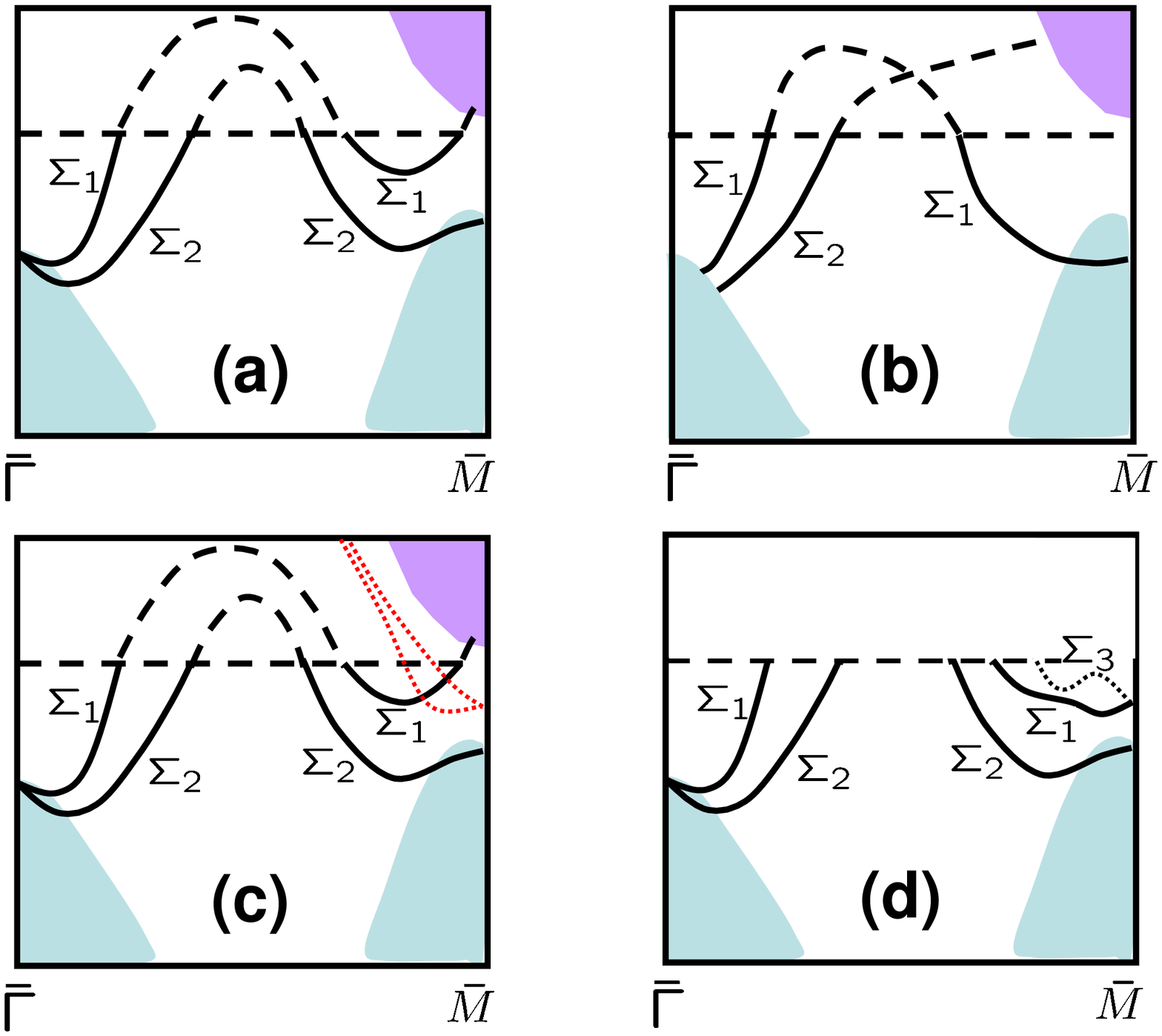}
\caption{Schematic picture for the comparison of the surface bands
  obtained from (a) our {\it ab-initio} calculation, (b) TB model
  (from the work of Teo {\it et al}\cite{teo2008}) and (d) ARPES experiment results. In
  ARPES experiment, an additional $\Sigma_3$ surface band (dotted line
  in (d)) becomes degenerate with $\Sigma_2$ band at $\bar{M}$ point.
  This additional band may come from the hybridization between the
  topological surface states and the other trivial surface states, as
  suggested by the red dotted line in (c).} \label{fig:exp}
\end{figure}

Here we compare our results with that from TB
analysis~\cite{fu2007B} and that from experiment~\cite{hsieh2008}.
As shown in Fig \ref{fig:exp} (a), we find five crossing points
between the surface bands and the Fermi energy along the line from
$\bar{\Gamma}$ to $\bar{M}$, this is the same to those observed in
the experiment of Hsieh {\it et al}\cite{hsieh2008} (Fig
\ref{fig:exp}), however in a simple TB model\cite{teo2008}, the
number of crossing is three (Fig \ref{fig:exp} (b)). A small
discrepancy is found between our {\it ab-initio} calculation and the
experiment of Hsieh {\it et al} near $\bar{M}$ point, as shown in
Fig \ref{fig:exp} (a) and (d). In the experiment of Hsieh {\it et
al}\cite{hsieh2008}, a third surface band $\Sigma_3$ appears near
$\bar{M}$ point and be degenerate with $\Sigma_1$ band at $\bar{M}$
point, however in our calculation, there is no such band and
$\Sigma_1$ band will go up and merge with the conduction band. This
discrepancy may come from additional trivial surface states, as
suggested in Fig \ref{fig:exp} (c). This discrepancy remains to be
justified by future studies.

\begin{figure}
\includegraphics[clip,width=7cm,angle=90]{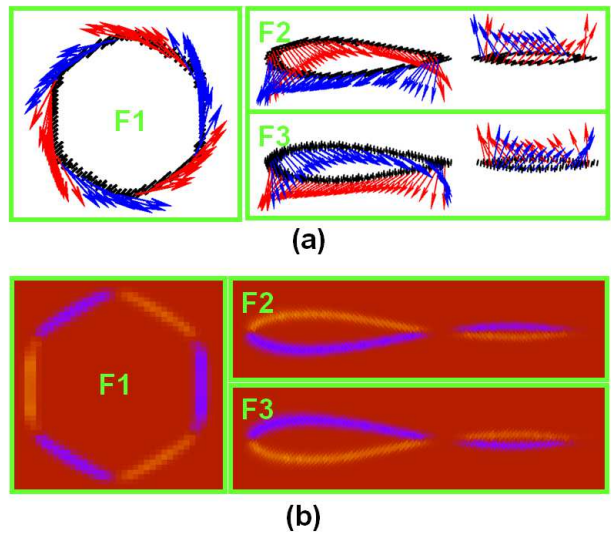}\\
\caption{(color online) The spin resolved Fermi surface for the
  semi-infinite Bi$_{1-x}$Sb$_{x}$'s below surface,
  whose normal is along -z direction. The arrow in
  (a) indicates $(S_x,S_y)$; different colors in (a) and (b)
  represent $S_z$ along different directions. The red color means that the $S_z$ is
  along the +z direction; the blue color means that the $S_z$ is along the -z direction.
  The  three pieces of Fermi surface $F_1$, $F_2$ and $F_3$ are marked
  in Fig \ref{fig:FStotal_topo}. $\lambda$ is taken as $1.1eV$
  here. } \label{fig:FSspin}
\end{figure}

In addition to the energy resolution, we are also able to calculate
the spin-resolved surface states. As an example, we carry out this
calculation for the semi-infinite Bi$_{1-x}$Sb$_{x}$ system's blow
surface, and show the spin-resolved surface state in
Fig.~\ref{fig:FSspin}.  The spin orientation of the surface states
at the Fermi level is plotted for three regions of 2D BZ with
$\lambda=1.1eV$. Clearly, vertex structure is found for the electron
pocket around $\Gamma$ point ($F_1$), which confirms the topological
nature of surface state. Because below surface's normal is along the
-z direction, we can confirm that the chirality of the vertex
structure around $\Gamma$ is left-handed and agrees with the recent
spin-resolved ARPES
experiment\cite{Hsiehspin2009}\cite{Nishidespin2009}.

\subsection{Indirect Band Gap}
\label{sec:Gap}

\begin{figure}
\includegraphics[clip,width=5.5cm,angle=-90]{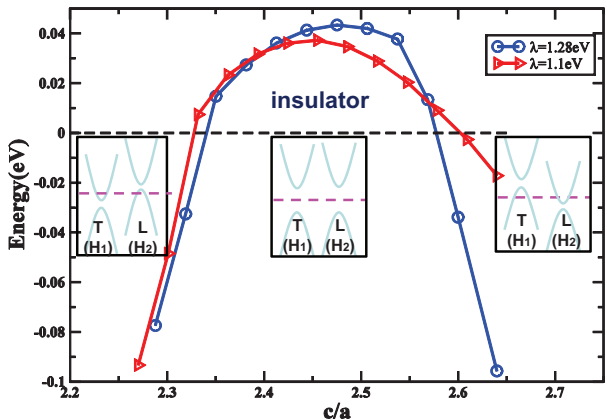}\\
\caption{(color online) Indirect energy band gap. We define the
indirect energy gap as the difference between conduction band
bottom(CBB) and the valence band top(VBT). The blue line and red
line present the indirect energy gap's functions of c/a for
$\lambda$=1.28eV and $\lambda$=1.1eV. H$_1$ and H$_2$ points are not
high symmetry points, and locate in the mirror plane in BZ. H$_1$ is
near to T point and H$_2$ is near to L point. In the inset, we show
the schematic figures indicating the different energy band
structures.} \label{fig:gap}
\end{figure}

In the above discussions, we call Bi$_{1-x}$Sb$_x$ bulk as
``insulator'' because there exists a direct band gap between the
conduction and the valence bands. Unfortunately, Bi$_{1-x}$Sb$_x$ is
actually a semi-metal (not true insulator) for most of the doping
range $x$, namely there exists finite overlap between the conduction
band bottom (CBB) and the valence band top (VBT). If we define the
true gap $E_g$ (indirect gap) as the energy difference between the CBB
and the VBT, $E_g$ is negative for most of the $x$, and it is positive
only for 0.07$<x<$0.22~\cite{teo2008}.  On the other hand, to identify
the TI nature, except the parity arguments as discussed above, it is
crucially important to have a full bulk gap throughout the
BZ. Therefore, a serious question for Bi$_{1-x}$Sb$_x$ is ``can we
make the indirect gap $E_g$ as positive as possible?'', or in other
words, ``can we widen the range of doping $x$ where system is truly
insulating?''. Here we will show that applying the uniaxial pressure
is an efficient way to open up the indirect band gap $E_g$.

Fig.11 shows the calculated indirect band gap $E_g$ as function of
$c/a$ ratio with fixed volume. The $c/a$ ratio can be tuned either
by $c$-axis pressure or by forming thin-film matched to substrate
with different lattice parameters. For both sides of the topological
phases ($\lambda$=1.28eV or 1.1eV), a broad positive $E_g$ region
can be obtained by reducing $c/a$ ratio slightly (around 3\%
reduction from its experimental value $c/a$=2.6). For both
$\lambda$=1.28eV and $\lambda$=1.1eV, the CBB is located at L point,
however the VBT is located at T point for $\lambda$=1.28eV, and at
H$_1$ point(around T point in the mirror plane) for $\lambda$=1.1eV.
Despite of the different positions of VBT, the effect of $c$-axis
pressure is always to raise the energy levels round L point, and
lower the levels around the T point. For $\lambda$=1.1eV, energy
level at H$_1$ point  goes lower, and energy level at H$_2$ point
near the L point in the mirror plane goes upper. Therefore the
positive indirect band gap is realized as schematically illustrated
in the insets of Fig.11.

\section{Conclusions}
\label{sec:conclusion}

As a summary, we develop a method to study the alloyed
Bi$_{1-x}$Sb$_x$ system and present a phase diagram to describe the
topological nature of the system. We show that Bi is topologically
trivial because the SOC is too strong. By alloying with Sb, the
effective SOC strength is reduced and the topologically non-trivial
phase is realized. By accurate {\it ab-initio} calculations and
MLWF, we calculate the surface states of semi-infinite system. The
results are compared with recent experiments. We predict the
spin-resolved Fermi surface which can be tested by spin resolved
ARPES. Finally, we suggest an efficient way to tune the indirect
band gap by uniaxial pressure, such that true bulk insulating state
can be realized for a broad doping range.

\section{Acknowledgments}
We acknowledge the support from Prof. N.~Marzari for solving
problems related to the construction of MLWF. This work is supported
by the National Science Foundation of China, the Knowledge
Innovation Project of the Chinese Academy of Sciences, and the 973
project of the Ministry of Science and Technology of China. XLQ and
SCZ are supported by the US Department of Energy, Office of Basic
Energy Sciences under contract DE-AC02-76SF00515.


%

\end{document}